# Direct visualization of domain wall pinning in sub-100nm 3D magnetic nanowires with cross-sectional curvature

*Joseph Askey‡, Matthew Oliver Hunt†‡, Lukas Payne, Arjen van den Berg, Ioannis Pitsios††, Alaa Hejazi†††, Wolfgang Langbein and Sam Ladak\**

School of Physics and Astronomy, Cardiff University, Cardiff, CF24 3AA, UK

Keywords: Two-photon lithography, direct laser writing, magnetic nanostructures, three-dimensional, domain walls, micromagnetics, magnetic force microscopy.

The study of 3D magnetic nanostructures has uncovered a range of rich phenomena including the stabilization and control of topological spin textures using nanoscale curvature, dynamic effects allowing controlled spin-wave emission, and novel ground states enabled by collective 3D frustrated interactions. From a technological perspective, 3D nanostructures offer routes to ultrahigh density data storage, massive interconnectivity within neuromorphic devices, as well as applications within health technologies, such as mechanical induction of stem cell differentiation. However, the fabrication of 3D nanomagnetic systems with feature sizes down to 10 nm poses a significant challenge. In this work we present a means of fabricating sub-100 nm 3D ferromagnetic nanowires, with both cross-sectional and longitudinal curvature, using two-photon lithography at a wavelength of 405 nm, combined with conventional deposition. Physical characterization illustrates that nanostructures with lateral features as low as 70 nm can be rapidly and reproducibly fabricated. A range of novel domain walls, with anti-vortex textures, coupled transverse textures, and hybrid vortex/anti-vortex textures are found to be enabled by the cross-sectional curvature of the system, as demonstrated by finite-element micromagnetic simulations. Magnetic force microscopy experiments in an externally applied magnetic field are used to image the injection and pinning of domain walls in the 3D magnetic nanowire. At specific field values, domain walls are observed to hop from trap to trap, providing a direct means to probe the local energy landscape. A simple model is presented demonstrating that thickness gradients and local roughness dictate the variation of pinning probability across the wire.

**Introduction**

Magnetic domain walls (DWs) in two-dimensional (2D) planar nanowire structures have been studied extensively for over 20 years providing platforms for applications in novel logic [1, 2], memory devices [3, 4], reservoir computing [5, 6], and biomedical applications [7]. More recently, studies have focused on fully three-dimensional (3D) nanowire structures, which provide access to novel spin textures [8-10], dynamic phenomena [9, 11, 12], and magnetochiral effects [13-15]. Alongside experimental progress, theoretical work has shown that nanoscale curvature can provide access to an effective Dzyaloshinskii-Moriya interaction (DMI) and an effective anisotropy interaction, providing additional degrees of control over spin textures [16-18]. Furthermore, it has been predicted that such curvature effects can pin DWs, confirmed by recent experimental work extracting the effective DMI for curved planar nanowires [19, 20].

A major technological driver for magnetic nanowires is the prospect of magnetic memory devices which rely on the motion and pinning of domain walls, a concept known as magnetic racetrack memory [3, 4]. This has been studied extensively in 2D nanostrips but ultimately requires 3D structures to enable densities competitive to V-NAND technology. Recent work by Gu et al. [21] demonstrated a multi-step methodology, combining electron-beam lithography and pattern transfer to realise suspended planar nanostrips on a modulated surface, creating a 3D racetrack memory device. This work represents a significant step forward in realizing 3D racetrack architectures.

Advancements in nano-fabrication technologies have enabled the wider realization of 3D ferromagnetic nanostructures by design. One such approach is focused electron beam induced deposition (FEBID) which uses a modified scanning electron microscope with a gas injection system [22]. This fabrication methodology has been used to realize a range of complex nanoscale magnetic architectures. Examples include scaffolds upon which Permalloy films have been thermally evaporated [23, 24], complex cylindrical magnetic nanowires coupled in a double-helix with feature sizes of order 80 nm and control over curvature [25, 26], as well as more complex nanostructures consisting of helices and connected nanowire networks [27, 28].

Another fabrication technique capable of realising 3D nanostructures is two-photon lithography (TPL), a form of direct laser writing (DLW). It can be combined with existing deposition techniques such as electrodeposition and thermal evaporation to realise 3D magnetic nanostructures [29-34]. Recent work has shown that these methods can produce 3D artificial spin ice (ASI) systems [33] which show controlled magnetic charge propagation [32], coherent spin-waves [35], and a rich phase diagram with a number of charge-ordered states [36]. Despite this success, a key disadvantage of commercial DLW systems is the limit on resolution and feature-size – typically about 200 nm in the lateral directions and 500 nm in the axial direction [29, 37]. This can be surmounted by combining DLW with a pyrolysis step, where polymer scaffolds are isotopically shrinking by slow decomposition during heating. This methodology has been reported by Pip et al. [38] where polymer tripods, with original features of 800 nm, were reduced by 70% from the original design to 160 nm. Another method of reducing feature sizes is to reduce the writing wavelength, utilising a 405 nm wavelength laser to achieve a polymeric feature-size below 100 nm [39].

Here we exploit DLW using a 405 nm wavelength laser to realize 3D magnetic nanowires with features as low as 70 nm, with both longitudinal and cross-sectional curvature. The novel crescent-shaped cross-section of the nanowire is found to stabilize a number of novel DW types, with the local curvature perturbing the spin texture, and in some cases, stabilizing topological defects such as anti-vortices. We demonstrate by magnetic force microscopy that such domain walls can be injected into the 3D nanowire and moved between pinning sites by using the field from the tip. A simple model based upon the local magnetostatic landscape of the wire is presented, suggesting that the observed DW pinning is due to the thickness gradient in sloped regions of the structure combined with local roughness.

**Results and Discussion**

An overview of the fabrication process (see Methods) is shown schematically in Figure 1. Resist is drop-cast onto a glass substrate (Figure 1a) into which a 3D geometry is exposed (Figure 1b). Development washes away unexposed regions yielding a freestanding 3D polymer (Figure 1c). Finally, thermal evaporation is used to coat Permalloy ($Ni_{81}Fe_{19}$) upon the 3D nanowires (Figure 1d). By harnessing a shorter $\lambda = 405$ nm laser to polymerize the resist, lateral features below 100 nm is possible [39].

The resulting structures are 3D ferromagnetic sinusoidal nanowires (SNWs) with both longitudinal and cross-sectional curvature. The longitudinal curvature of the SNWs is controlled through the spatial wavelength $L$ of the sinusoidal exposure trajectory with values $L$ = 1 µm, 2 µm and 5 µm, whilst its amplitude is fixed at $A = 500$ nm, as sketched in Figure 2(a). The mean lateral feature sizes $w_{avg}$ of the fabricated SNWs for the three spatial periods were measured using SEM (see Methods) and are plotted in Figure 2(b) as a function of inverse scanning velocity $1/v$ which is proportional to the effective exposure dose. The data points correspond to the measurements and the solid lines to a fitted profile of the form $w_{avg} = 2b \sqrt{\ln(v_{th}/v)}$ (modified from reference [40] assuming a Gaussian dose shape), where $v$ is the scanning velocity, $v_{th}$ is the threshold velocity and $b$ is the width of the dose shape (see Methods). We find that SNWs with feature sizes as low as 70 nm can be readily fabricated. The fits generally yield threshold velocities and voxel widths which agree within the estimated measurement error (see Methods). Example SEM images of SNWs with nominal lateral widths of 80 nm are shown in Figure 2(c) for $L = 1$ µm, Figure 2(d) for $L = 2$ µm and Figure 2(e) for $L$ = 5 μm. In addition to SEM, the lateral features of SNWs with spatial period $L$ = 5 μm were further measured using quantitative differential interference contrast (qDIC) microscopy prior to thermal evaporation (see Methods) [41-44]. This provides a direct measurement of the polymer cross-sectional area, and by assuming the cross-section is an ellipse, we can determine the axial length of the polymer and thus the aspect ratio of the voxel volume (see Supplementary Figure S1). Analysis of the qDIC images (see Methods) yields a lateral width $w$ = 65 nm and axial length $l$ = 160 nm close to the value for a numerically simulated diffraction limited focal spot (see Figure 3a). The measured qDIC and SEM widths agree to within 10% (see Methods).

Magnetic nanowires fabricated using the combination of two-photon DLW and thermal evaporation are known to result in a crescent-shaped cross-section [31] where similar structures have recently been studied numerically, elucidating high-frequency spin-wave dispersion relations [45]. However, the DW textures that these systems may yield have not yet been studied in detail. Finite-element micromagnetic simulations (see Methods) were used to determine the spin texture of domain walls that one might expect in the experimental SNWs studied in this work. The SNWs host two distinct geometric curvatures: the longitudinal curvature $\kappa_l$, defined by the spatial period of the sinusoid trajectory, Figure 2(a); and the cross-sectional curvature $\kappa_c$, defined by the shape of the voxel, Figures 3(a)-(c). To ensure an accurate geometric mesh in the simulations, the focused laser point spread function (PSF) was calculated (see Methods, Supplementary Figure S2) and is shown in Figure 3(a), yielding a voxel width $l_{xy}$ = 66 nm, and voxel height $l_z$ = 164 nm, and an aspect ratio $\beta$ = 2.48 at 80% total squared intensity, parameters which are close to the physical characterization.

The Permalloy deposition process was then simulated by linear extrusion of the calculated PSF along the wire, as shown schematically in Figure 3(b). Previous work has shown that such thermal evaporation of NiFe normal to the substrate yields a crescent-shaped cross-section as depicted in Figure 3(b) lower-panel. The geometry of the magnetic nanowire cross-section is defined as the difference of two ellipses, $a$ and $b$, with major radii $s_a = 120$ nm and $s_b = 80$ nm, and minor radii $r_a = 42$ nm and $r_b = 40$ nm, Figure 3(c). The peak cross-sectional curvature value $\varkappa_c = 18.8$ μm$^{-1}$ which yields $\kappa_c \approx 8\kappa_l$ for the $L = 1$ μm case. The peak longitudinal curvatures in the experimental systems are $\varkappa_l = 0.10$ μm$^{-1}$ for $L = 5$ μm; $\varkappa_l = 0.62$ μm$^{-1}$ for $L = 2$ μm; and $\varkappa_l = 2.47$ μm$^{-1}$ for $L = 1$ μm. Therefore, one would expect that any curvature-induced perturbation of the DW textures would arise from $\varkappa_c$ rather than $\varkappa_l$, and thus we model here straight segments of wire with purely cross-sectional curvature. Figures 3(d-g) shows the ground state domain wall configuration comprising two transverse-like textures with aligned in-plane components, to yield a continuous magnetization transition across the wire apex (Figure 3d) without a topological defect. The two sides of the DW are then found to have opposing $m_z$ as shown in Figures 3(e-f), minimizing the magnetostatic energy. A 3D representation of the wall shows a canting of magnetization into the $z$ direction due to the curvature-induced DMI. We now briefly consider the higher energy DWs obtained in simulations. Supplementary Figure S3 shows the stabilization of an anti-vortex wall (AVW) which features two transverse-like textures but both with $-m_z$ component. Since the magnetization on either side of the curvature peak are anti-aligned (Figure S3a), together with the opposing magnetization along the wire axis, this provides the boundary condition for an anti-vortex (topological charge, $n=-1$) at the curvature peak. Defects of total topological charge $n=1/2$ are located on each opposing wire edge. We note that such anti-vortex walls have not been observed in conventional planar nanowires. The walls are likely enabled by the 3D curved geometry allowing both sides of the nanowire to have the same $m_z$ component, whilst being spatially separated (reducing the magnetostatic energy), The geometry also yields an effective DMI strongest at peak curvature, stabilizing the central portion of the anti-vortex which is aligned in the *z*-direction. Compared with the ground state CTW, the AVW has approximately 2% larger total energy, mainly due to the texture surrounding the anti-vortex and the out-of-plane component at the anti-vortex core. Even more complex domain wall types appearing at higher energy (see Supplementary Table 1) are shown in Supplementary Figures S4 and S5. Overall, the micromagnetic simulations demonstrate a range of novel domain wall spin textures which occur due to the 3D wire geometry with strong cross-sectional curvature.

With a good understanding of the possible domain walls, we proceeded to carry out imaging experiments aimed at injection and subsequent pinning of DWs in the 3D nanowire, using magnetic force microscopy (MFM) under externally applied magnetic fields (see Methods). An example 5 μm × 5 μm AFM image of a SNW with period $L = 5$ μm is shown in Figure 4(a), and a 3D representation of the same SNW is given in Figure 4(b). The SNWs studied have a width around 80 nm and are fabricated close to the polymerization threshold, yielding a strong dependence of voxel size on exposure dose. This enhances the surface and edge roughness due to the temporal fluctuations in the laser power during writing, and allows pinning of walls at specific wire locations. The fast scan axis in Figure 4(a) is aligned with the long axis of the SNW, the *x*-axis. The slow scan axis (i.e motion of the tip along the *y*-axis) is disabled once the image acquisition is shown to be scanning along the SNW, as indicated by the red dashed line in Figure 4(a). By scanning along this direction, the stray field from the tip $\mathbf{H}_{tip}$, when superposed with an external in-plane field $\mathbf{H}_{ext}$, allows the DW to explore its local energy landscape and pinning events are observed directly. In Figure 4(c-d), a fixed field $\mu_0|\mathbf{H}_{ext}|= 0.2$ mT is applied along the positive *x*-direction for the entire image capture and the image is captured top-to-bottom. The data show stochastic pinning and de-pinning of DWs on the SNW. To gain insight into pinning positions along the SNWs, multiple MFM images are taken for different fixed fields, and the DW positions are identified on a line-by-line basis (see Methods), as shown in Figure 4(e) where the red dots correspond to the DW positions. Figure 4(f) illustrates the normalized counts of pinning events observed in Figure 4(e). To build a statistical representation of the pinning potential across the wire, the observed DW positions are collated into histograms by binning over a 16-pixel range. We note that this analysis was not meaningful for $L = 1$ μm SNWs since the observed DW contrast was exclusively located at the edges of the SNW – a hypothesis for this behavior is discussed later in the manuscript.

SEM images taken at a 45° viewing angle for $L = 5$ μm and $L = 2$ μm are shown in Figures 5(a-b), respectively, with corresponding DW position heatmaps in Figures 5(c-d). For both wires, one can see that even for a fixed value of $\mu_0|\mathbf{H}_{ext}|$ the DW positions can fluctuate between different pinning sites along the SNW axis, as is especially the case for $\mu_0|\mathbf{H}_{ext}| = 0.2$ mT for $L = 5$ μm (Figure 5c), and $\mu_0|\mathbf{H}_{ext}| = 0.5$ mT for $L = 2$ μm (Figure 5d). This indicates that the DW pinning and motion along the SNWs possess some degree of stochasticity, likely driven by the stray field from the MFM tip during the scans, though this is somewhat reduced where $\mu_0|\mathbf{H}_{ext}| = 0$ mT in both the $L = 5$ μm and $L = 2$ μm cases. There are also clear and distinct bands of DW pinning over large ranges in the $-\mu_0|\mathbf{H}_{ext}|$ for the $L = 5$ μm case at $x = 0.78$ μm, $x = 2.34$ μm and $x = 3.28$ μm, which indicates that there is a significant pinning potential that must be overcome in these regions. Such bands of DW pinning are not as significant in the $L = 2$ μm case in either $\pm\mu_0|\mathbf{H}_{ext}|$ but there are bands of weaker pinning at $x = 0.12$ μm and $x = 0.88$ μm in the $+\mu_0|\mathbf{H}_{ext}|$ panel.

The peaks of the SNWs are approximately 1.72 μm for $L = 5$ μm, and 1.00 μm for $L = 2$ μm. For $L = 5$ μm there is no significant pinning at the apex, with strong bands of pinning occurring in regions along the ascending regions of the SNW. Weak pinning is observed at the apex of the $L = 2$ μm case, though these pinning events mainly fall into the $x = 0.88$ μm bin rather than the $x = 1.22$ μm, likely driven by the positive field direction which propagates the DW in the $+x$-direction, though the reverse situation for a negative field does not result in stronger pinning in the $x = 1.22$ μm bin. The lack of, or weakly observed, pinning at the regions of maximal longitudinal curvature contrasts with similar experimental and theoretical studies on curvature induced DW pinning in ferromagnetic parabolic stripes, where DW pinning was observed at local curvature maxima [46, 47]. Recent theoretical work on similar parabolic stripes with tailored cross-sectional thicknesses has illustrated that spatially varying cross-sections can provide additional sources of geometry-induced DMI and anisotropy [48], and can lead to DW pinning regions adjacent to the typical pinning sites at the local curvature maxima where the cross-section is at a minimum. In the present case, the spatially varying cross-section, Figure 3(c), is not believed to induce the pinning observed in Figures 5(b) and 5(g) as the cross-section is approximately constant along the SNW period, so we consider other reasons for the observed pinning landscape.

DW pinning typically arises due to the presence of material defects which can include dislocations, impurities, grain boundaries and other inhomogeneities; polycrystallinity and surface roughness [49]. It is well-known that thermally evaporated $Ni_{81}Fe_{19}$ in identical systems used in the set-up at present produce highly pure deposits with ratios of Ni to Fe being 4.26 [31, 33], with small grains of less than a few nm, has a well-defined face-centered cubic crystal structure in films thinner than 60 nm [50] and has negligible magnetocrystalline anisotropy. We therefore consider SNW surface roughness induced pinning, with clear roughness features observable in the SEM images in Figures 5(a-b), as the potential source of the observed behavior in Figures 5(c-d). AFM line profiles of the $L = 5$ μm and $L = 2$ μm SNWs are shown in Figures 5(e-f), respectively. We first consider the total field $\mathbf{H}_{tot}$ experienced by decomposing the external field $\mathbf{H}_{ext}$ and the field from the tip $\mathbf{H}_{tip}$ along the local SNW tangent (see Methods), where the absolute values $\mu_0|\mathbf{H}_{tot}|$ are shown in Figures 5(g-h) for $L = 5$ μm and $L = 2$ μm, respectively. The blue curves correspond to the maximal external field $\mu_0|\mathbf{H}_{ext}|_{max}$, and the red curves to the minimum external field $\mu_0|\mathbf{H}_{ext}|_{min}$, with the green shaded regions corresponding

to the intermediate values investigated in the partial hysteresis loops. Both Figure 5(g) and 5(h) reveal the rather strong field component provided by the tip (5-10 mT), which facilitates DW movement in regions where the nanowire tangent has an out-of-plane component.

Following the work of Schöbitz et al. [49] and Bruno [51], we derive an expression for the depinning field due to roughness, arising from magnetostatic contributions to anisotropy, using a Becker-Kondorski model [52]. The result of this analysis (see Methods) predicts a depinning field $H_{BK} = \frac{9}{160} \frac{M_s \sigma}{t}$, where $\sigma$ is the surface root-mean-square (RMS) roughness, and $t$ is the thickness of the $Ni_{81}Fe_{19}$ layer. We therefore expect the depinning field to scale with roughness and inversely with thickness and consider both in our analysis. Simple geometric considerations yield a local thickness $t' = t \cos\theta$ where $\theta$ is the angle between nanowire tangent and substrate and $t$ is the thickness deposited on a flat region. The local RMS roughness for $L$ = 5 μm with value $\sigma$ = 7.6 nm can vary up to 60% over the nanowire length, and for $L$ = 2 μm with value $\sigma$ = 8.1 nm which can vary up to 75% over the nanowire length, whilst local variations in wire thickness are up to 15% for $L$ = 5 μm and up to 50% for $L$ = 2 μm. It is therefore anticipated that the sites of largest depinning field will arise from the combined effects of reduced thickness and increased roughness, with roughness dominating for larger $L$, and thickness variations dominating for smaller $L$ (see Supplementary Figure S6). The depinning field $\mu_0|\mathbf{H}_{BK}|$ for both geometries is calculated directly using the AFM topography within the Becker-Kondorski framework and is shown in Figures 5(i-j) for the $L$ = 5 μm and $L$ = 2 μm SNWs, respectively. Three cases are plotted, varying roughness and constant thickness; constant roughness (equal to the RMS roughness for each nanowire) and varying thickness (green); and the combination of varying roughness and thickness (red). It can be seen that the depinning fields for both wires are the same order of magnitude as the applied fields, suggesting pinning events are likely to be observed. Two regimes of pinning can be identified in both nanowires, with higher depinning fields observed in regions of high slope due to the reduced thickness, and lower depinning fields observed at flatter regions near the center, driven by the local roughness, largely matching the general pinning trend seen in Figures 5(c-d). Differences in the experimental pinning probabilities between the two nanowires can be understood by considering the field component tangential to the wire direction. The $L = 2$ μm nanowire, though possessing a higher calculated depinning field in regions of higher slope, also has a larger out-of-plane field component along the wire, making strong pinning less likely and instead the DW hops between pinning sites in sloped regions. In contrast, the $L = 5$ μm nanowire is seen to have well-defined bands of pinning with probability close to 1. Here, the calculated depinning field is of similar magnitude to that of the $L = 2$ μm but the component of the field along the nanowire is reduced, favouring pinning of DWs at a single site. For $L = 1$ μm (see Supplementary Figure S6) the depinning fields were found to be $\mu_0|\mathbf{H}_{BK}| = 90$ mT at one edge, whilst the total field along the SNW tangent never exceeded $\mu_0|\mathbf{H}_{tot}| = 20$ mT, lending insight into the lack of stochastic DW pinning.

Overall, the model provides a spatial variation of depinning field that is generally consistent with the experimental data, when also considering the total magnetic field at different points on the wires. We note that some discrepancies between the calculated pinning and experimental data are observed. Most notable is at $x = 1.88$ μm, for the $L = 2$ μm wire where $|\mathbf{H}_{tot}| \ll |\mathbf{H}_{BK}|$, and yet no pinning behavior is observed. Such discrepancies could be related to the following three points. Firstly, the total field $\mu_0|\mathbf{H}_{tot}|$ is estimated assuming a perfectly aligned tip perpendicular to the sample surface. Slight deviations due to substrate tilt or any imperfections in tip fabrication will yield perturbations in the field experienced. Secondly, the calculated depinning fields are based on a model with planar approximation and this considers neither the non-planar cross-section of the nanowires nor the 3D stray field of the underlying domain walls. Thirdly, we note that the forward and backward scanning of the tip, when combined with a geometric variation in pinning potential will yield stochasticity in DW movement.

## Conclusions

We have presented a novel means to fabricate sub-100 nm 3D ferromagnetic nanowires with cross-sectional and longitudinal curvature using a modified two-photon lithography direct laser writing system. Physical characterization of these 3D nanostructures shows that we can consistently fabricate nanostructures with lateral features below 100 nm and axial features below 245 nm, as predicted by the numerically modelled PSF. Finite-element micromagnetic simulations demonstrate that such nanowires with crescent-shaped cross-section are home to several exotic DWs not previously reported in the literature. MFM in externally applied magnetic fields has been used to study nucleation and pinning of DWs in the nanowires. The observed pinning behavior can be understood within the context of thickness gradients and surface roughness across the nanowire. It is expected that careful tuning of exposure settings and geometric parameters of the fabrication will allow a reduction in surface roughness, whilst a more conformal coating of magnetic material may be achieved by using atomic layer deposition, as demonstrated recently on 3D woodpile structures [60], or using substrate tilt and rotation during conventional deposition. Ultimately, our method can be used to realize 3D magnetic nanostructures of chosen geometry, by design, allowing the study of a wide range of topological spin textures and their time-dependent phenomena.

## Methods

**Direct laser writing.** The nanowire fabrication is performed using a modified two-photon lithography system (Photonic Professional GT, Nanoscribe GmbH). A continuous-wave diode laser of wavelength $\lambda = 405$ nm with a peak power of 100 mW (Photon Lines, Omicron, LDM405) is used as the excitation beam. The output beam is passed through an optical isolator (Crystalaser, I-405-03) to suppress back-reflections into the diode cavity. The laser power is controlled and digitally modulated by an acousto-optic modulator (AA acousto optics, MQ110-A3-UV) driven by a fixed frequency driver (AA acousto optics, MODA110-D4-36). The digital modulation input of the driver is fed by an arbitrary function generator (RSPro, AFG-21005), providing a square wave of 1 MHz frequency and 20% duty cycle. A plano-convex lens (Thorlabs, LA1131A, $f = 50$ mm) focuses the first-order Bragg diffraction of the AOM through a 50 $\mu$m pinhole, blocking other orders. The divergent beam is reflected by a dielectric mirror (EKSMA, reflection > 99.5% for $\lambda = 380 – 420$ nm) and collimated using an achromat (Edmund Optics #49362, $f = 150$ mm). The mirror also transmits the light reflected by the sample, forming an image of the sample on a CMOS camera (Basler ACE CMOS, acA2040-120μm), allowing the determination of the relative position of the reflected focus image with respect to the glass-resist interface. The collimated beam is then focused by an achromat (Edmund Optics #49358, $f = 75$ mm) onto the intermediate image plane at the left port of the microscope stand, enters the port, is deflected upwards by the port prism, collimated through the tube lens (Zeiss, $f = 160$ mm), and converted from linearly to circularly polarized light using a quarter-wave plate (Thorlabs WPQ05M405). The beam is focused into the sample plane by a 63× 1.4NA objective lens (Zeiss, Plan-Apochromat 420782-9900-799) using oil-immersion ($n = 1.518$, Zeiss Immersol 518F). The Gaussian excitation beam diameter at $1/e^2$ intensity entering the objective is 1.67 times the objective pupil diameter. A two-photon susceptible photoresist at $\lambda = 405$ nm (IP-Dip NPI, Nanoscribe GmbH) is drop-cast onto a square glass coverslip (22 mm × 22mm, 0.16 – 0.19 mm) and is exposed with a fixed average transmitted laser power $P = 2.8$ mW. The sinusoidal exposure trajectory is controlled via piezoelectric stages in which the sample is mounted. The exposure dose is varied by the stage scanning velocity increasing in 3% steps from an initial velocity $v_0 = 45$ μm/s up to a maximum velocity $v_{max} = 109$ μm/s. Following the exposure, the sample is developed in propylene-glycol-methyl-ether-acetate (PGMEA) for 20 minutes to remove the unexposed photoresist, followed by isopropanol (IPA) for 5 minutes. The polymer sinusoidal nanowires are then coated top-down with a 40 nm layer of Permalloy ($Ni_{81}Fe_{19}$) using thermal evaporation as follows. A ribbon of $Ni_{81}Fe_{19}$ is washed in isopropanol and mounted in an alumina-coated molybdenum boat, and the system is pumped down to a pressure of $10^{-6}$ mbar. The thermal evaporation results in open-shell ferromagnetic nanowire structures (see Figure 3) with both longitudinal curvature due to the sinusoidal trajectory and cross-sectional curvature due to the asymmetry in lateral and axial size of the point-spread function (PSF) of the laser focus, and the directional $Ni_{81}Fe_{19}$ thermal evaporation.

**PSF Calculation.** The electric field of the focused light through a high-NA objective lens is calculated numerically using vectorial diffraction theory evaluating of the Debye-Wolf integral using a chirp-z transform (or Bluestein method), following the work by Hu et al. [53] and Leutenegger et al. [54]. The $\lambda$ = 405 nm circularly polarized voxel focused through an oil-immersion ($n$ = 1.518) objective lens with NA = 1.4 is shown in Figure 3(a), illustrated by the intensity square ($I^2$) isosurfaces (since the exposure dose $\propto I^2$), with the innermost and darkest isosurface corresponding to 80% of the total normalized $I^2$ (sequentially larger and lighter isosurfaces correspond to 60, 40 and 20%). The voxel width is $l_{xy}$ = 66 nm and the voxel length $l_z$ = 164 nm, leading to aspect ratio $\beta$ = 2.48. The FWHM of the lateral $I^2$ profile is 118 nm. From the measured widths in Figure 2, one can also determine the FWHM of the excitation voxel assuming a Gaussian excitation pulse with intensity of the form $I(r) = \frac{1}{v}\exp\left(-(r/b)^2\right)$ where $v$ is the scanning velocity, $r$ is radial distance from the beam focus, and $b$ (the fitting parameter) is the lateral width of the voxel, using FWHM = $2\sqrt{\ln(2)}\,b$, which leads to a FWHM = 107 nm for $L$ = 5 μm , which is a 9% difference compared to the simulated FWHM.

For $L = 2$ μm we find FWHM = 90 nm yielding a 24% difference, and for $L = 1$ μm a FWHM = 113 nm giving a 4% difference to the simulation.

**Micromagnetics.** The finite-element micromagnetic software NMag [55] is used to simulate magnetization textures in a planar nanowire with cross-sectional curvature only, see Figure 3(b). The magnetization dynamics are governed by the Landau-Lifshitz-Gilbert equation composed of a damping and precession term: $d\mathbf{m}/dt = (-\gamma_0/1 + \alpha^2)[\mathbf{m} \times \mathbf{H}_{\text{eff}} + \alpha \mathbf{m} \times (\mathbf{m} \times \mathbf{H}_{\text{eff}})]$, with the reduced magnetization vector $\mathbf{m} = \mathbf{M}/M_s$ using the saturation magnetization $M_s$. The gyromagnetic ratio is $\gamma_0 = 2.211 \times 10^5$ mA$^{-1}$s$^{-1}$ and we have chosen a Gilbert damping parameter of $\alpha = 0.5$ for all simulations to reduce computation time without impact on the final relaxed state [56]. The effective field $\mathbf{H}_{\text{eff}} = (-1/\mu_0 M_s)\partial \varepsilon_{\text{tot}}/\partial \mathbf{m}$ has a total energy density $\varepsilon_{\text{tot}}$ comprised of an exchange and magnetostatic component $\varepsilon_{\text{tot}} = \varepsilon_{\text{exc}} + \varepsilon_{\text{mag}}$. The wires are simulated with Ni$_{81}$Fe$_{19}$ parameters with $M_s = 8.6 \times 10^5$ Am$^{-1}$ and exchange stiffness A$_{\text{ex}} = 1.3 \times 10^{-11}$ Jm$^{-1}$, leading to an exchange length $l_{\text{ex}} = \sqrt{2A_{\text{ex}}/\mu_0 M_s^2} = 5.3$ nm. The geometry is meshed using NetGen [57] with a minimum node spacing of 2 nm and a maximum spacing of 3.5 nm. A wire length of 1 μm was used in all micromagnetic simulations. The central 300 nm segment is initialized with a random magnetization, and the magnetization in the lateral segments are fixed pointing towards or away from the central randomized region, forming head-to-head (H-H) or tail-to-tail (T-T) DWs upon relaxation, respectively. The relaxation simulations have been repeated 20 times, with different random seeds, for both H-H and T-T configurations. The simulations are stopped when $|d\mathbf{m}/dt| < 1.745 \times 10^{-2}$ deg s$^{-1}$, where convergence is satisfied when the largest value of $|d\mathbf{m}/dt|$ drops below this value.

**Characterisation.** SEM was performed using a Hitachi SU8230 in a vacuum with pressure of $1 \times 10^{-4}$ mbar, 3 kV or 5 kV accelerating voltages and 10 μA probe current. The fits shown in Figure 2(b) give threshold velocities of $v_{th}$ = (133 ± 6) μm/s and lateral voxel width $b$ = (64 ± 3) nm, for $L$ = 5 μm; $v_{th}$ = (179 ± 14) μm/s and lateral voxel width $b$ = (54 ± 3) nm, for $L$ = 2 μm; and $v_{th}$ = (143 ± 9) μm/s and lateral voxel width $b$ = (68 ± 3) nm, for $L$ = 1 μm. qDIC was performed on a custom-built imaging set-up as described by Regan et al. [42] and Hamilton et al. [58]. The imaging was carried out using a Nikon green interference filter (center wavelength $\lambda$ = 550 nm), a de-Sénarmont compensator (a rotatable linear polariser and quarter-wave plate, Nikon T-P2 DIC Polariser HT MEN51941) controlling the phase offset, an oil immersion 1.34NA condenser MEL41410 with a Nikon N2 DIC module MEH52500, and a water-immersion ($n$ = 1.333) 60× 1.27 NA objective lens (Nikon plan-apochromat MRD70650) with a DIC slider (Nikon MBH76264), a linear polariser (Nikon Ti-A-E DIC Analyser Block MEN51980), and a 1× tube lens. All images were acquired using a scientific-CMOS camera (PCO Edge 5.5 RS, PCO) of full well capacity of 30 ke and 16-bit digitisation. Pairs of DIC images using exposure time of 2.7 ms and a field of view size of 555 μm × 468 μm (2560 × 2160 pixels) were taken at polarizer angles of ±30°, $I_{\pm}$ and combined into a contrast image using $I = (I_+ - I_-)/(I_+ + I_-)$, and then converted into a qDIC image via the procedure detailed in [41]. The cross-sectional area $A_c$ and polymer width $w$ are calculated as the average of two signal-to-noise ratio qDIC images, $\varkappa$ = 500 and $\varkappa$ = 5000, with values $A_c$ = 32.59 μm² and $w$ = 65 nm. This leads to a mean axial extent $l = \frac{A_c}{\pi w} = 160$ nm, for exposure parameters $P_q$ = 1.76 mW and $v_q$ = 64 μm/s with exposure dose $D_q \propto \frac{P_q^2}{v_q} = 0.6 D_s$ compared with SEM parameters of $P_s$ = 2.80 mW and $v_s$ = 100 μm/s. By extrapolating the fits made in Figure 2(b) for $L$ = 5 μm and using an adjusted SEM scanning velocity $v_s = 0.6 v_q \frac{P_s^2}{P_q^2} = 97$ μm/s, one finds that an equivalent lateral feature size $w \approx 72$ nm, with percentage difference of 10% with respect to the qDIC measured value. This discrepancy can be attributed to the finite resolution in the qDIC measurement, and the signal-to-noise ratios used in the analysis of the dataset. The aspect ratio of the cross-section $\beta = l/w = 2.44$, is close to the expected voxel aspect ratio for a diffraction limited focal spot $\beta = 2.46$ (see Figure 3), with a percentage difference of less than 1%. AFM and MFM (Bruker Dimension 3100) was performed using low-moment super sharp MFM tips (Nanosensors SSS-MFMR) with radius of curvature less than 15 nm, and magnetic moment $|\mu| \approx 2.5 \times 10^{-17}$ Am². All AFM and MFM images were processed using Gwyddion analysis software [59], where the images are plane levelled and tip-strike artefacts removed wherever possible. In all MFM images the lift-height was set to a constant $z = 70$. The MFM images are first binarized, and each line of the image is then background subtracted and fitted with a Gaussian function where the center position of the Gaussian is taken as the position of the DW. The field from the tip is estimated using a dipolar expression $\mu_0 \mathbf{H}_{\text{tip}} = \frac{\mu_0}{4\pi} (3 \mathbf{r} (\mathbf{\mu} \cdot \mathbf{r})/|\mathbf{r}|^5 - \mathbf{\mu}/|\mathbf{r}|^3)$, with magnetic moment $\mathbf{\mu} = (0, 0, \mu_{\text{tip}})$, and position vector $\mathbf{r} = (0, 0, z)$, i.e the magnetic moment and the position vector are assumed to be purely a function of $z$. This estimation leads to a value of $\mu_0|\mathbf{H}_{\text{tip}}| = 14.6$ mT. The MFM images were captured with a variable externally applied magnetic field $\mu_0|\mathbf{H}_{\text{ext}}|$, using a bespoke quadrupole electromagnet where the applied field was oriented along the wire long axis, labelled $x$-axes in Figures 4 and 5. The total field $\mu_0|\mathbf{H}_{\text{tot}}|$ is the projection of the external field $\mu_0|\mathbf{H}_{\text{ext}}|$ and the tip field $\mu_0|\mathbf{H}_{\text{tip}}|$ along the local SNW tangent $|\mathbf{H}_{\text{tot}}| = |\mathbf{H}_{\text{ext}}| \cos(\phi) + |\mathbf{H}_{\text{tip}}| \cos(\theta)$ where $\phi = \tan^{-1}(dz/dx)$ and $\theta$ are the angles subtended between the $z$-axis (in and out of plane) and the $x$-axis (along longitudinal direction of SNW), respectively. The height derivative of the AFM profiles $dz/dx$ was smoothed using a 3-pixel rolling average, corresponding to a spatial averaging of 30 nm.

**Estimating Depinning Fields**. The Becker-Kondorski model predicts that local minima in the energy landscape gives rise to DW pinning, and that the depinning field $H_{BK}$ is proportional to the slope of the position-dependent energy landscape $\varepsilon(x)$ [49, 52], which can be written as $|\mathbf{H}_{BK}| = (1/2\mu_0 M_s S) d\varepsilon/dx$, where $S$ is the cross-sectional area. We simplify the analysis by considering a simple planar strip with comparable geometric parameters with width $w = 80$ nm and thickness $t = 40$ nm, where $S = wt$, and the surface area over element $\delta x$ is $\mathcal{S} = w\delta x$. To estimate $|\mathbf{H}_{BK}|$ we simplify the energy $\varepsilon(x)$ formulation by Bruno et al. by considering that roughness features along the SNW are not correlated (as we attribute power fluctuations in the laser give rise to the dominant roughness along the SNW), such that we can estimate $\varepsilon(x) = 0.45\, \mu_0 \mathcal{S} M_s^2 \frac{\sigma}{4}$, where $\sigma$ is the RMS roughness of the SNW in an element $\delta x$. The slope of this potential landscape is therefore $\frac{d\varepsilon}{dx} = 0.45 \mu_0 w M_s^2 \frac{\sigma}{4}$, and the depinning field in the present case is $|\mathbf{H}_{BK}| = \frac{9}{160} \frac{M_s \sigma}{t}$. The RMS roughness $\sigma$ is determined by extracting the high frequency components of the AFM profile using a cut-off frequency 20% of the total addressable length in the profile. This corresponds to 720 nm for $L = 5$ μm; 430 nm for $L = 2$ μm, and 420 nm for $L = 1$ μm. The roughness as a function of position $\sigma(x)$ is then calculated over the binning sizes of the DW pinning heatmaps shown in Figure 5 with value $\delta x = 313$ nm for $L = 5$ μm, and $\delta x = 250$ nm for $L = 2$ μm. For $L = 1$ μm, the roughness $\sigma$ is calculated over an element equal to the cut-off frequency length with value $\delta x = 420$ nm.

AUTHOR INFORMATION

**Corresponding Author**

LadakS@cardiff.ac.uk

**Present Addresses**

†Matthew Hunt, Huntleigh Healthcare Ltd, Cardiff, U.K.

††Ioannis Pitsios, VitreaLab GmbH, Vienna, Austria.

††Alaa Hejazi, Taibah University - Faculty of Science and Arts, Janada Bin Umayyah Road, Medina 42353, Kingdom of Saudi Arabia

**Author Contributions**

The manuscript was written through contributions of all authors. All authors have given approval to the final version of the manuscript. ‡These authors contributed equally. (match statement to author names with a symbol).

Conceptualization, S.L. and W.L.,.; Methodology, M.H., J.A., L.P., I.P., S.L., W.L., ; Software, M.H. and J.A.; Validation, M.H., J.A., L.P., I.P., A.V. D.B.; Formal Analysis, M.H. and J.A.; Investigation, M.H., J.A., L.P., I.P.; Resources, S.L. and W.L.; Data Curation, M.H. and J.A.; Writing – Original Draft Preparation, M.H., J.A. and S.L.; Writing – Review & Editing, M.H., J.A., S.L., W.L.; Visualization, J.A.; Supervision, S.L. and W.L.; Project Administration, S.L. and W.L.; Funding Acquisition, S.L. and W.L.

ACKNOWLEDGMENT

This work was funded by the Engineering and Physics Research Council (EPSRC) grant numbers EP/R009147/1 and EP/X012735/1. S.L. also acknowledges funding from the Leverhulme Trust, grant number RPG-2021-139.

ABBREVIATIONS

DW (Domain Walls); DMI (Dzyaloshinskii-Moriya Interaction); FEBID (Focused Electron Beam Induced Deposition); TPL (Two-Photon Lithography); DLW (Direct Laser Writing); ASI (Artificial Spin Ice); SEM (Scanning Electron Microscopy); AFM (Atomic Force Microscopy); MFM (Magnetic Force Microscopy); SNW (Sinusoidal Nanowire); qDIC (Quantitative Differential Interference Contrast); CTW (Coupled Transverse Wall); AVW (Anti-Vortex Wall).

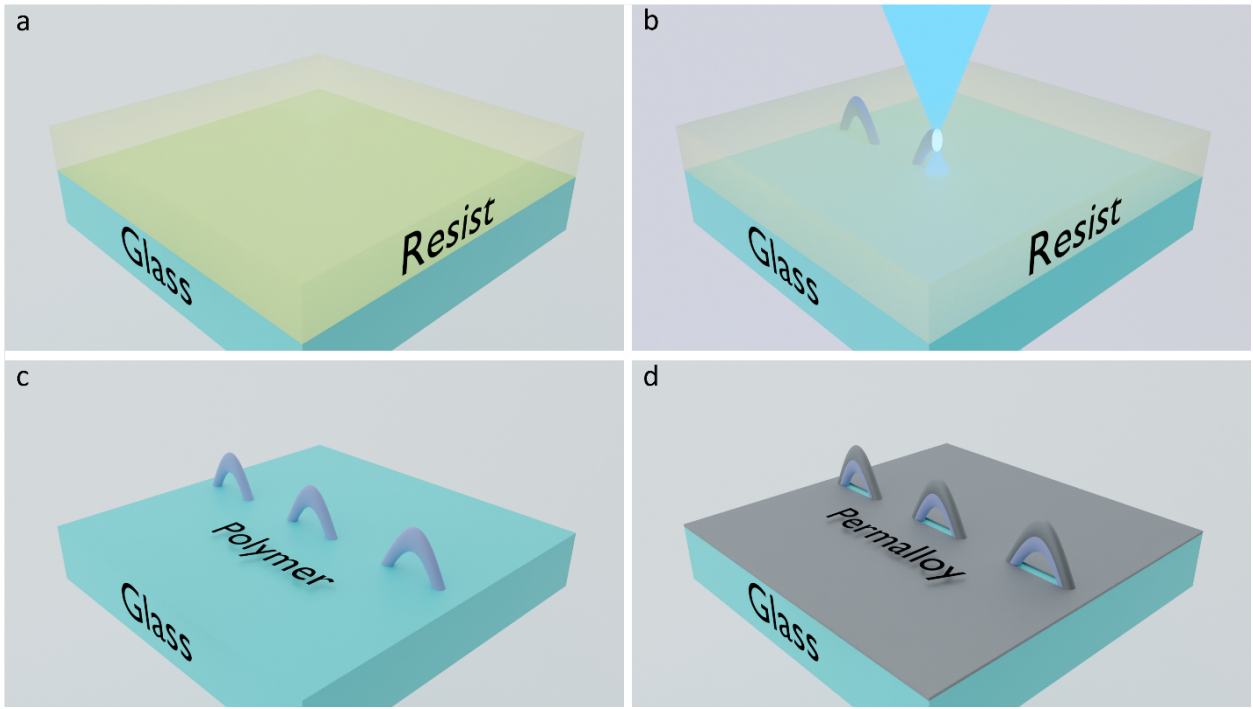

**Figure 1: Overview of the fabrication process.** (a) Negative-tone photoresist (yellow) drop-cast on glass substrate (blue). (b) Exposure of wavelength $\lambda$ = 405 nm laser. (c) Unexposed resist is removed through solvent development and polymer (purple) nanowires remain. (d) Thermal evaporation of Permalloy ($Ni_{81}Fe_{19}$, grey) is used to coat the polymer nanowires.

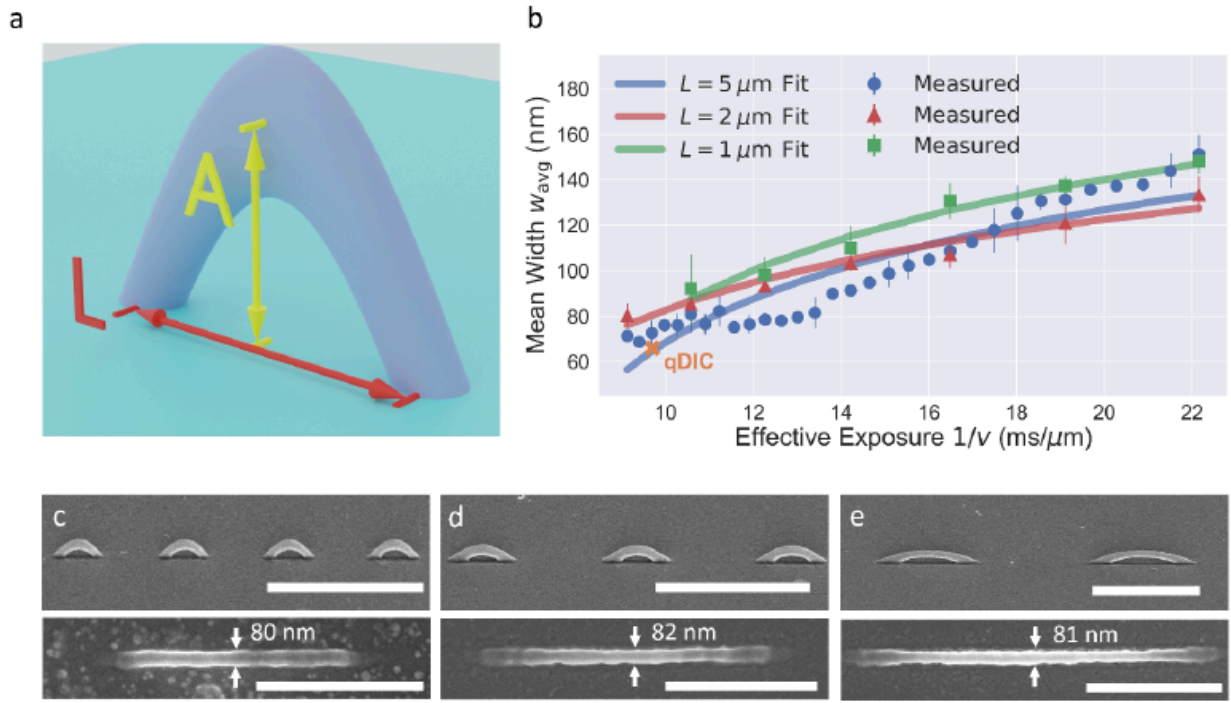

**Figure 2**: **Physical characterization of 3D magnetic nanowires**. (a) Schematic of a SNW with spatial period $L$ (red) and amplitude $A$ (yellow). (b) Average SNW width $w_{avg}$ for $L = 5$ µm (blue circles), $L = 2$ µm (red triangles) and $L = 1$ µm (green squares) measured using SEM. Solid lines show fits to expected dependence. The orange cross shows the width as determined by qDIC before evaporation. (c) $L = 1$ µm, (d) $L = 2$ µm and (e) $L = 5$ µm SEM images taken from 45° (top panel), where wires are fabricated well above the polymerization threshold, and top-down (bottom-panel) views, where wires are fabricated close to polymerization threshold. Variations in substrate tilt lead to wires with smaller experimental length. Scale bars are 5 µm and 1 µm in top and bottom panels, respectively.

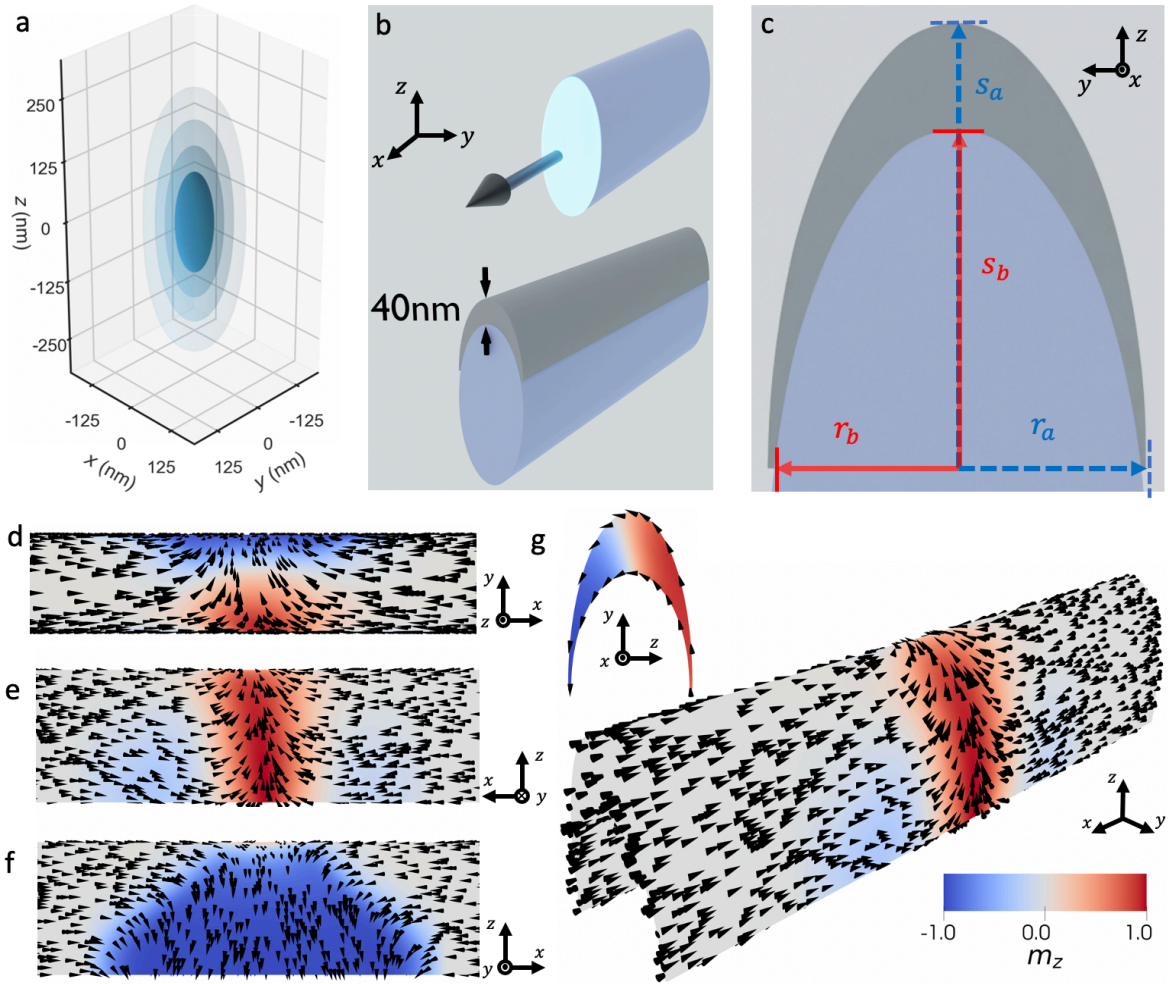

**Figure 3: Harnessing the point spread function curvature to realise novel domain walls**

(a) Voxel of a circularly polarized λ = 405 nm laser focused through a NA = 1.4, 63× oil immersion (n = 1.518) objective lens. The innermost (dark blue) isosurface corresponds to a polymerization threshold of 80% the total I². The voxel width $l_{xy}$ = 66 nm, and voxel height $l_z$ = 164 nm, and aspect ratio β = 2.48. (b) Illustration of a voxel (bright blue) scanning uniformly in a lateral direction creating a polymer (purple) wire (top), and 40 nm $Ni_{81}Fe_{19}$ (grey) evaporation on top of this wire (bottom). (c) Cross-sectional view of the nanowire mesh, with outer major and minor radii $s_a$ and $r_a$ (blue dashed arrows) and inner major and minor radii $s_b$ and $r_b$ (red solid arrows). (d) Coupled-transverse domain wall (CTW) top-down (top panel), (e,f) side-view and (g) 3D view with inset showing cross-section. Color scheme corresponds to $m_z$ component of the reduced magnetization vector **m**.

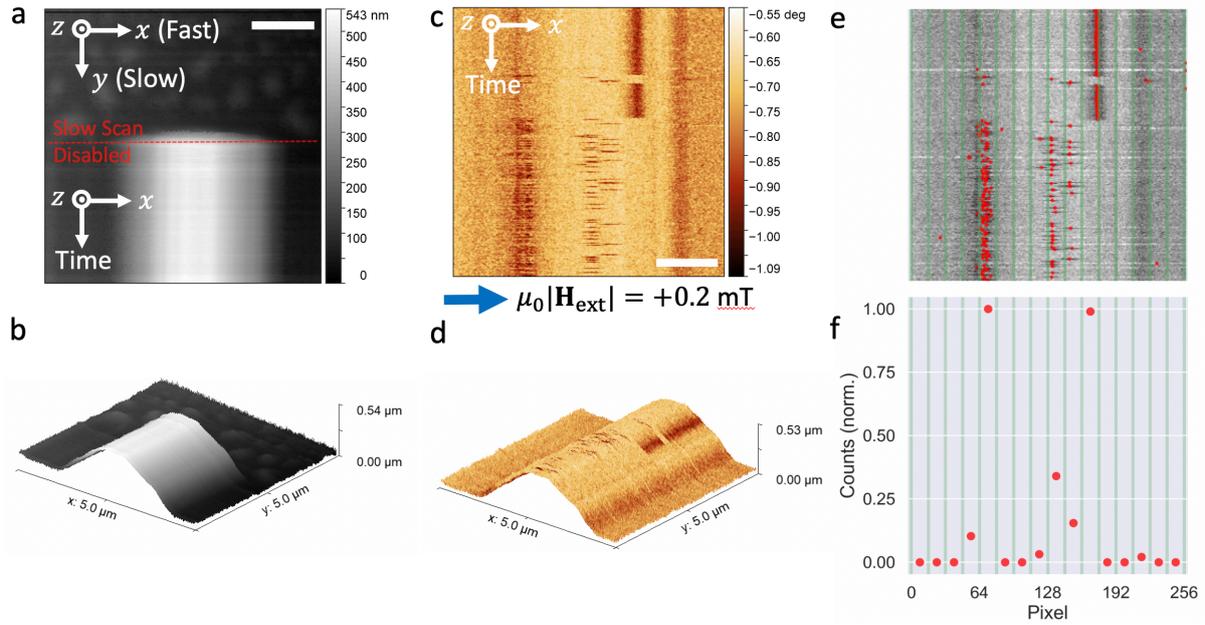

**Figure 4. Magnetic force microscopy procedure.** (a) Example AFM image of $L = 5$ μm SNW, with fast and slow scanning axes indicated by white arrows. The slow scan axis is paused when scanning along the SNW. (b) 3D view of the AFM image. (c) MFM image with fixed positive external field $\mu_0|\mathbf{H}_{ext}| = +0.2$ mT (blue arrow) applied along the $+x$-direction. Image capture is top-to-bottom. A head-to-head DW is stochastically pinned and de-pinned along the wire during the scan. (d) 3D view of the MFM image in (c), illustrating the positions of the domain wall. (e) Fitted Gaussian peak positions (red dots) corresponding to domain wall positions along the SNW. (f) Normalised counts of domain wall positions as function of pixel index of (e), green lines correspond to a 16-pixel binning window. Scale bars in all images are 1 μm.

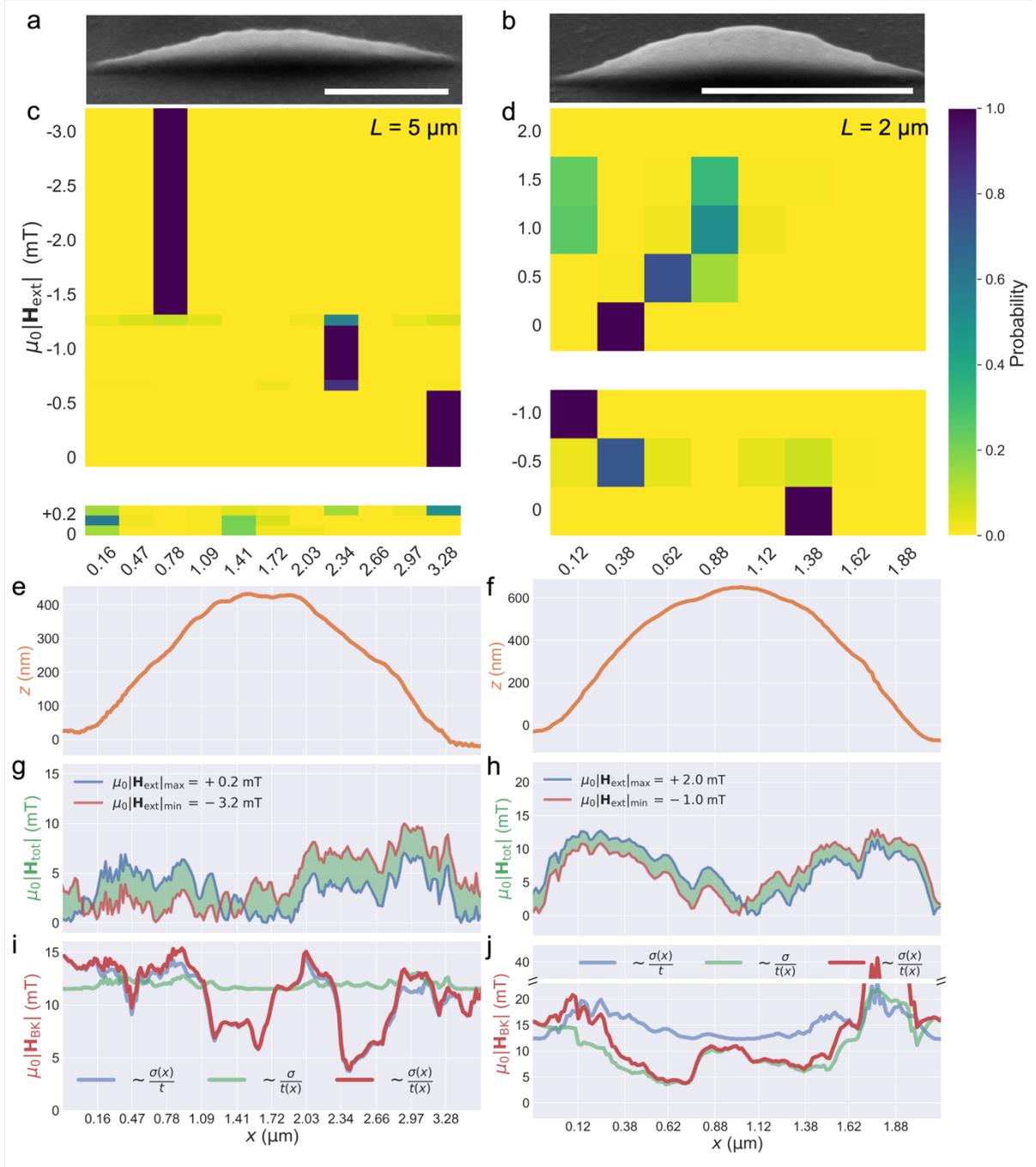

**Figure 5. Probing the energy landscape of 3D magnetic nanowire.** (a) and (b) SEM images taken at 45° viewing angle of L = 5 μm and L = 2 μm SNWs, respectively, scale bars 1 μm. (c) and (d) DW pinning position probability maps as function of longitudinal position $x$ along the SNW and the applied field $\mu_0|\mathbf{H}_{ext}|$. (e) and (f) AFM line profiles used in the determination of the SNW roughness. (g) and (h) total field magnitude $\mu_0|\mathbf{H}_{tot}|$ due to external field $\mu_0|\mathbf{H}_{ext}|$ and field from MFM tip $\mu_0|\mathbf{H}_{tip}|$ projected along the SNW tangent. (i) and (j) Becker-Kondorski depinning field magnitude $\mu_0|\mathbf{H}_{BK}|$ with varying roughness constant thickness (blue line),

constant roughness varying thickness (green line), and varying roughness varying thickness (red line).

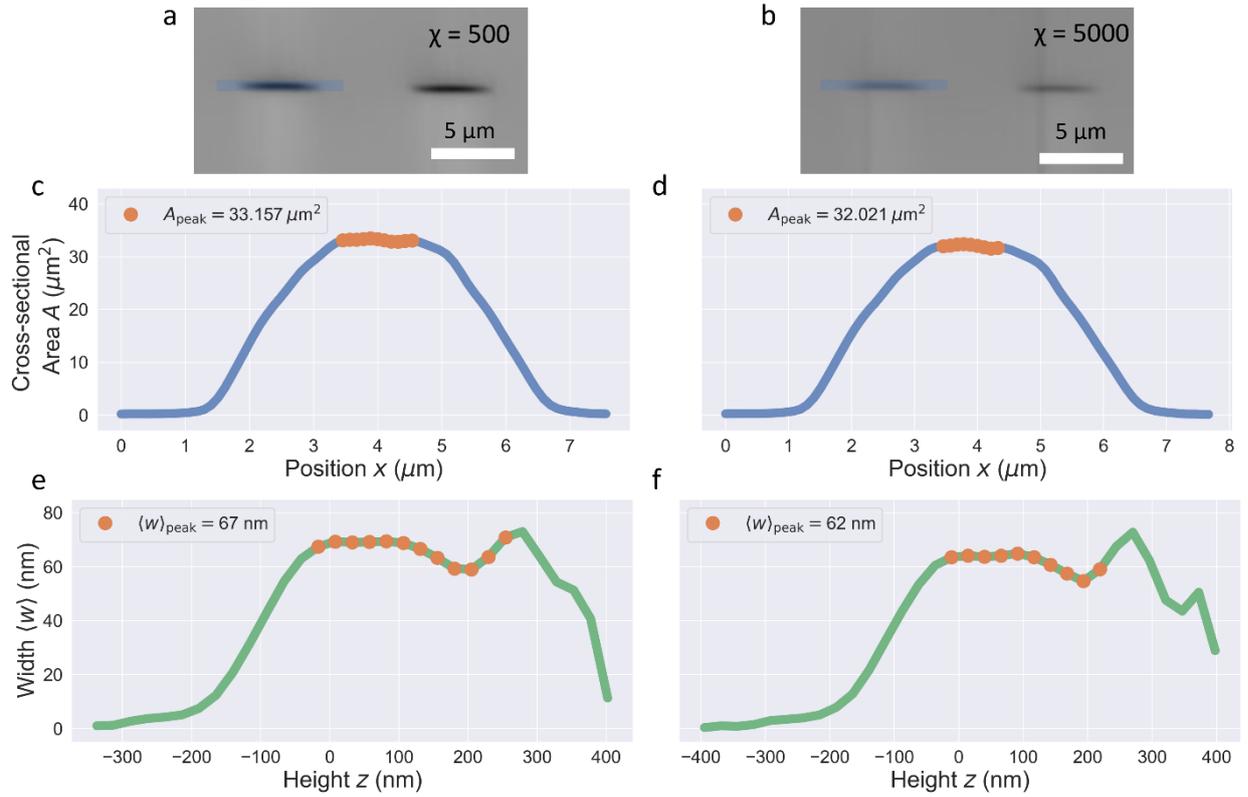

**Figure S1.** (a) Quantitative differential interference contrast phase images of $L = 5$ μm and $A = 1$ μm. SNWs with different signal-to-noise ratios for $\chi = 500$, (b) and $\chi = 5000$. Scale bar is $L = 5$ μm. (c) Extracted SNW cross-sectional areas $A$ as function of lateral position (as indicated by shaded blue lines in the phase images) for $\chi = 500$, (d) and $\chi = 5000$. The orange data points correspond to the regions where the SNW is completely extruded above substrate, and the peak cross-sectional area $A_{peak}$ is taken as the mean of these values. (e) Mean width $\langle w \rangle$ of ascending and descending regions of the SNW as function of SNW height above substrate for $\chi = 500$, (f) and $\chi = 5000$. The mean peak width $\langle w \rangle_{peak}$ is taken as the mean of the orange data points.

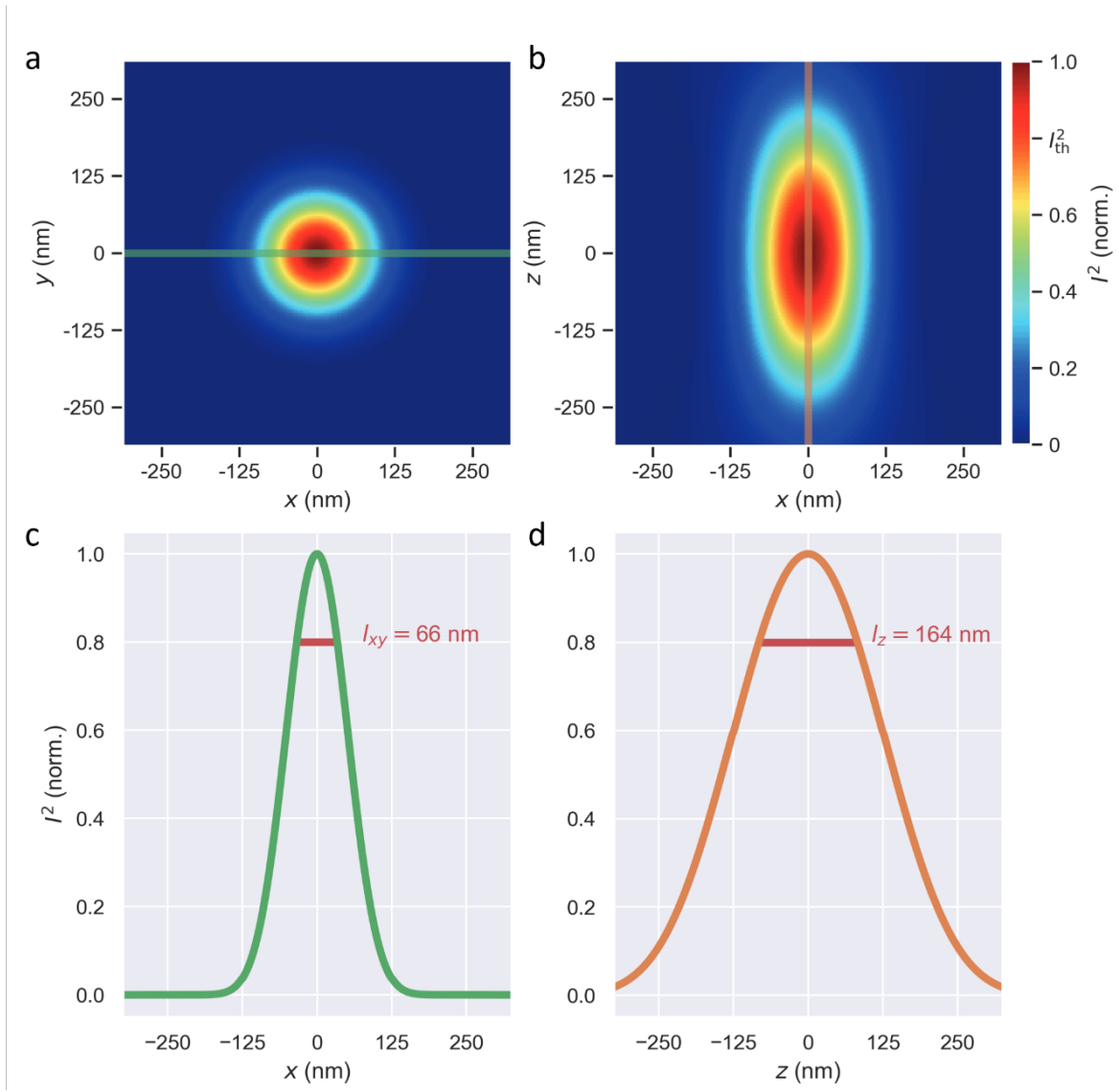

**Figure S2.** (a) Numerically simulated lateral intensity square $I^2$ profile of circularly polarized $\lambda = 405$ nm beam focused through an NA $= 1.4$ objective lens into an immersion medium with refractive index $n = 1.518$ at $z = 0$, fill factor 1.67. (b) Axial profile at $y = 0$. (c) Lateral line profile of the focus, green line in (a), with $l_{xy} = 66$ nm at threshold ($I^2 = 0.8$). (d) Axial line profile of focus, orange line in (b), with $l_z = 164$ nm at threshold.

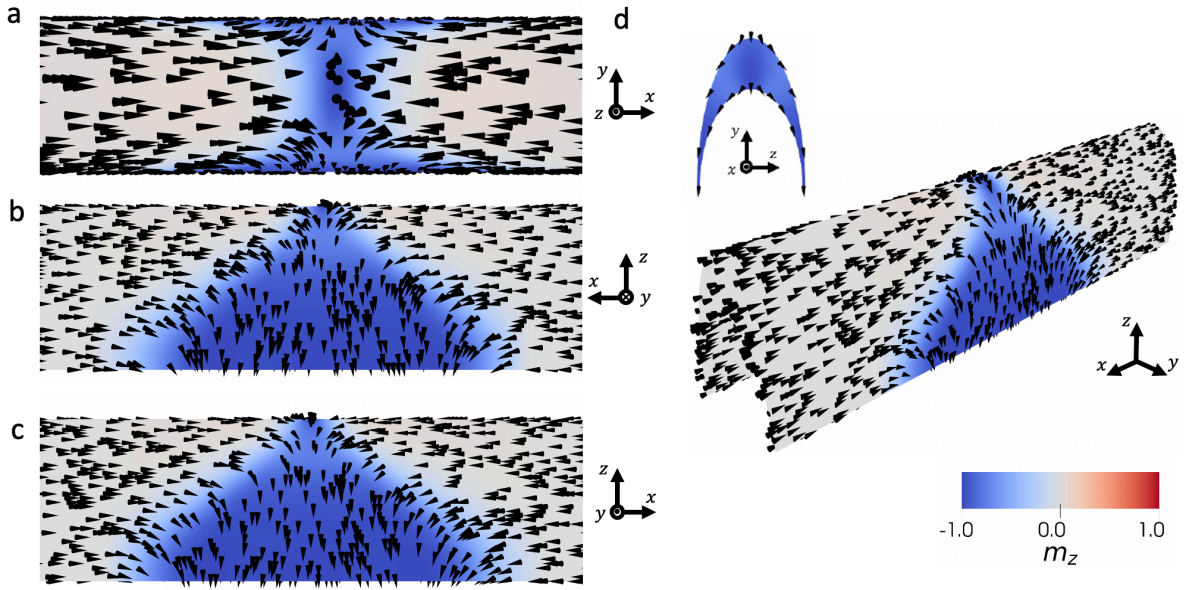

**Figure S3:** An anti-vortex domain wall texture. (a) Top view. (b,c) Side views. (d) 3D view. The wall consists of a transverse spin texture on either side, with an anti-vortex stabilized at the curvature apex.

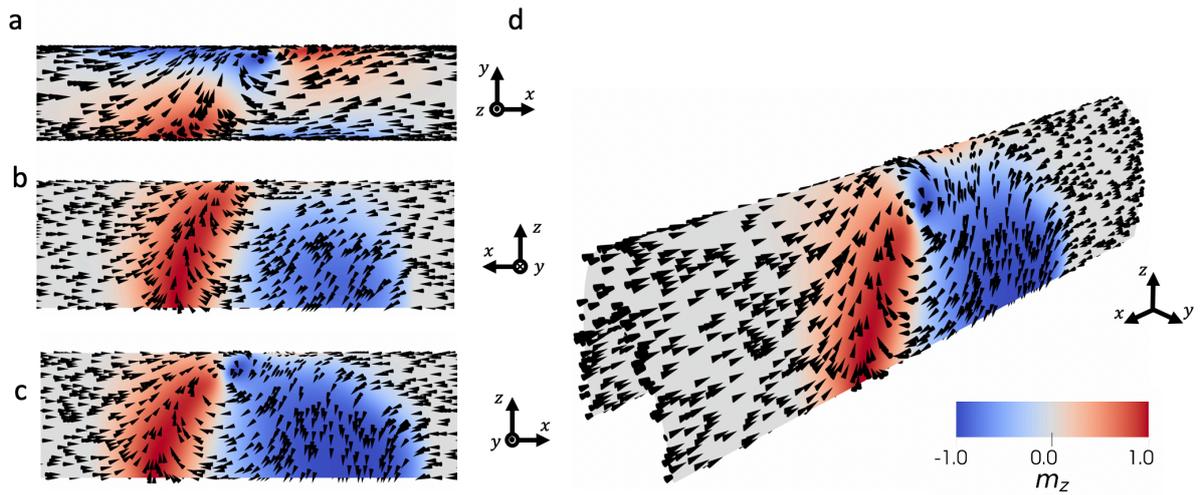

**Figure S4:** A vortex domain wall texture. (a) Top view. (b,c) Side views. (d) 3D view. The wall consists of a single vortex texture, that spans across both sides of the nanowire. The vortex core is found to be located just off the apex of curvature.

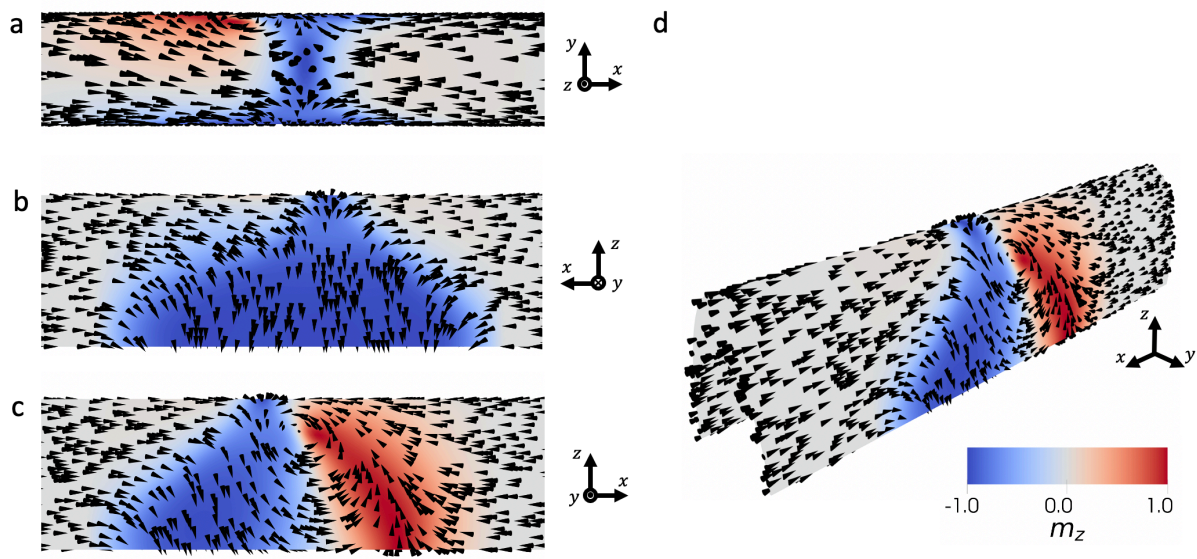

**Figure S5:** An Anti-vortex/vortex domain wall texture. (a) Top view. (b,c) Side views. (d) 3D view. The wall consists of a transverse spin texture on one side, with vortex on the remaining side. An anti-vortex is found to be stabilized at the apex of curvature.

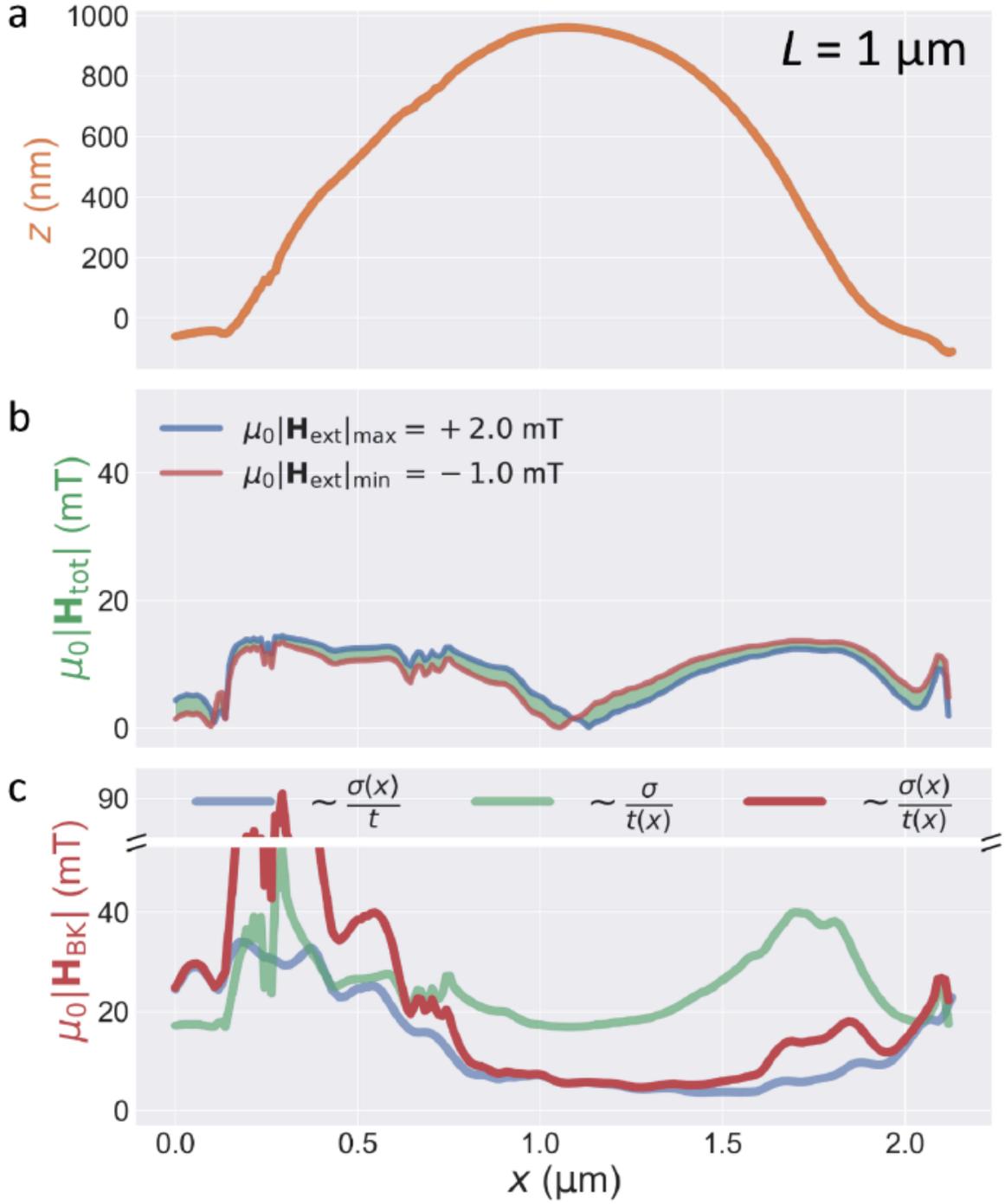

**Figure S6** (a) AFM line profile for a L = 1 μm SNW. (b) Total field magnitude $\mu_0|\mathbf{H}_{tot}|$ due to external field $\mu_0|\mathbf{H}_{ext}|$ and field from MFM tip $\mu_0|\mathbf{H}_{tip}|$ projected along SNW tangent. (c) Becker-Kondorski depinning field magnitude $\mu_0|\mathbf{H}_{BK}|$ for varying roughness constant thickness (blue line), constant roughness varying thickness (green line) and varying roughness and thickness (red).

| DW Type | $\varepsilon_{exc}$ (J/m$^3$) | | $\varepsilon_{mag}$ (J/m$^3$) | | $\varepsilon_{tot}$ (J/m$^3$) | |
|---|---|---|---|---|---|---|
| | **H-H** | **T-T** | **H-H** | **T-T** | **H-H** | **T-T** |
| **CTW** | 1672.601 | 1671.421 | 11484.55 | 11485.59 | 13157.15 | 13157.01 |
| **AVW** | 1714.26 | 1714.173 | 11686.9 | 11687.03 | 13401.16 | 13401.2 |
| **TVW** | 3630.671 | 3632.208 | 9891.096 | 10819.29 | 13521.77 | 13521.27 |
| **AVVW** | 2933.599 | 2935.199 | 10820.09 | 9889.062 | 13753.69 | 13754.49 |

**Table 1:** Energy density components for all simulated domain wall types in both head-to-head (H-H) and tail-to-tail (T-T) configurations.